\documentclass[10pt,preprint]{aastex}  

\slugcomment{To appear in the Astrophysical Journal, Part 1, Nov 10, 2003}

\begin{document}  

\title{A Study of Cyg OB2: Pointing the Way Towards Finding\\
Our Galaxy's Super Star Clusters}
\author{M.M. Hanson$^1$}
\affil{Department of Physics, The University of Cincinnati, Cincinnati, 
OH 45221-0011}
\altaffiltext{1}{Visiting astronomer, Steward Observatory, University of Arizona}

\begin{abstract}
New optical MK classification spectra have been obtained for 
14 OB star candidates identified by Comer\'{o}n et al.\ 
(2002) and presumed to be possible members of 
the Cyg OB2 cluster as recently described by Kn\"{o}dlseder 
(2000). All 14 candidate OB stars observed 
are indeed early-type stars, strongly suggesting the 
remaining 31 candidates by Comer\'{o}n et al.\ are also 
early-type stars.  A thorough investigation of the properties of
these new
candidate members compared to the properties of the Cyg OB2
cluster star have been completed, using traditional as well
as newly revised effective temperature scales for O stars.  
The cooler O-star, effective temperature scale
of Martins et al. (2002) gives a very close distance for the 
cluster (DM = 10.4).  However, even using traditional effective 
temperature 
scales, Cyg OB2 appears to be slightly closer (DM = 10.8) than 
previous studies determined (DM = 11.2; Massey \& Thompson 1992), 
when the very young age of the stellar cluster ($\sim$ 2 x 10$^6$ yrs)
is taken into account in fitting the late-O and early-B dwarfs
to model isochrones.  
Of the 14 new OB stars observed for this study, as many 
as half appear to be significantly older than the 
previously studied optical cluster, making 
their membership in Cyg OB2 suspect. So, while some of the 
newly identified OB stars may represent a more extended halo 
of the Cyg OB2 cluster, the survey of Comer\'{o}n et al.\ 
also picked up a large fraction of non-members. Presently, 
estimates of the very high mass of this cluster ($M_{cl} 
\approx 10^4 M_{\odot}$ and over 100 O stars) first made by 
Kn\"{o}dlseder (2000) remain higher than this study can support. 
Despite this, the recognition of Cyg OB2 being a more 
massive and extensive star cluster than previously realized 
using 2MASS images, along with the recently recognized
candidate super 
star cluster Westerlund~1 only a few kpc away (Clark \& 
Negueruela 2002), reminds us that we are 
woefully under-informed about the massive cluster 
population in our Galaxy. Extrapolations of the locally
derived cluster luminosity function indicate 10s to perhaps
100 of these very massive open clusters ($M_{cl} \approx 10^4 
M_{\odot}$, $M_V \approx -11$) should exist within our 
galaxy. Radio surveys will not detect 
these massive clusters if they are more than a few million 
years old. Our best hope for remedying this shortfall is 
through deep infrared searches and follow up near-infrared 
spectroscopic observations, as was used by Comeron et al.\
(2002) to identify 
candidate members of the Cyg OB2 association.
\end{abstract} 


\section{INTRODUCTION}  

An accurate census of our Galaxy's massive star content is 
essential to understanding its evolution and structure
(see Russeil 2003).  
Yet until very recently, few open clusters were known 
to exist with extinction greater than A$_V > 5$. 
Moving our imaging and spectroscopic studies to 
the near-infrared, specifically around 1 to 2~$\mu$m, 
has begun to uncover significant massive star populations
previously unknown or poorly understood.  
Because of its conspicuous
nature, the galactic center region was among the first to
be deeply studied in the near-infrared.  These studies identified 
at least three spectacular and very massive clusters (Morris \& 
Serabyn 1996; Figer et al.\ 1999; Eckart, Ott \& Genzel 1999)
with unique stellar and cluster properties.  Arguments 
about the unusual nature of the galactic center environment have
 been used to explain the galactic center cluster characteristics
(Morris \& Serabyn 1996), but how unique are the galactic center 
clusters to other massive clusters lurking through out our galaxy?  
Might similar very massive clusters lie in other regions of our 
galaxy?  

The recently completed 2MASS survey (Skrutskie et al.\ 1997)
represents a critical 
first step in uncovering massive open clusters
in our galaxy.  Not only are numerous new clusters being found 
through near-infrared surveys (Dutra \&  Bica 2000; 2001; 
Ivanov et al. 2002), but even well known clusters are
being found to be more extensive and massive than previously 
thought.  Such is the case for
Cygnus OB2, a massive star cluster less than 2 kpc away.
Originally discovered and labeled IV Cygni by Munch \& Morgan (1953),
Cyg OB2 has been the focus of numerous optical studies, dating
back nearly 50 years (Johnson \& Morgan 1954; Morgan et al.\ 1954; 
Schulte 1956, 1958).  The most thorough investigation of Cyg~OB2 
was completed by Massey \& Thompson (1991, hereafter MT91), 
identifying 120 possible massive star members based on UBV photometry, 
and giving optical spectral classifications for just over 70 OB stars in 
the field.  Recently, Kn\"{o}dlseder (2000), using 2MASS infrared 
observations, has reevaluated the stellar content of Cyg OB2, arguing 
for a more massive and extensive cluster than previously recognized. 
Kn\"{o}dlseder (2000) proposed Cyg OB2 to be the most massive 
stellar association known in our Galaxy, and referred to it as
a ``super star cluster'', containing over 100 O stars 
(Kn\"{o}dlseder et al.\ 2002).

The Kn\"{o}dlseder (2000) study of Cyg OB2 was completed 
using near-infrared imaging photometry alone.  
Though far more time consuming, and not possible for very heavily 
reddened sight-lines, optical spectroscopic studies, such as that
completed by MT91 on Cyg OB2, give us the needed 
information to confidently characterize a cluster's mass, age,
initial mass function (IMF), and to accurately study the individual members.
As a follow up to the Kn\"{o}dlseder (2000) study, Comer\'{o}n et 
al.~(2002, here after CPR2002) presented a low resolution $H$- and $K$-band 
spectroscopic survey of 2MASS identified sources towards Cyg OB2.  
While these spectra were not of the
quality required for near-infrared spectral classification (see Hanson,
Howarth and Conti 1997), they were sufficient to quickly confirm
which stars already showing ``blue'' near-infrared colors lacked
discernible molecular bands, consistent with them being early-type
stars.  They identified 77 early-type candidates. 
Less than half, just 31 stars, had been previously classified with
optical spectra to have been OB stars, leaving 46 new candidate OB
stars towards Cyg OB2.

In this study, we have obtained classification-quality blue spectra 
for 14 of the 46 OB star candidates identified by CPR2002 to 
determine their MK classifications.  A 15th star from the CPR2003 list
was observed, but it turned out to have been previously studied 
spectroscopically.  The primary goal of this study
is to determine if the OB candidates identified using 2MASS colors and
low-resolution near-infrared survey spectra are indeed OB stars. 
In this way, the observations presented here serve as a useful test 
of this newly explored, but clearly important, near-infrared method 
of identifying OB star populations through out our galaxy and behind 
large line-of-sight extinction.  Observations are presented in \S 2,
and the new spectra are presented in \S 3.

A second goal of this paper, which is more difficult than its first, 
is to determine if the newly found OB stars are associated with the 
optically studied Cyg OB2 cluster of MT91.  This will require a thorough
evaluation of the cluster's characteristics to determine the likelihood 
of membership for any newly found OB stars in the region.  A review
of the Cyg OB2 cluster characteristics and the characteristics of the
newly found OB stars is given in \S 4.  A final discussion of the
search for possible super star clusters within our Galaxy is presented
in \S 5.  Concluding remarks are found in \S 6.

\section{OBSERVATIONS AND REDUCTIONS}

Observations were made the nights of 6, 7, 8 July, 2002 on the University 
of Arizona's 2.3m Bok Telescope, located on Kitt Peak, outside of Tucson, AZ.
The Boller and Chivens (B\&C) spectrograph was employed and operated with an 
832 g/mm grating and a Schott 8612 order separating filter.  A full-width, 
half-maximum resolution of FWHM $\approx$ 2.0 \AA\ (about 2.8 pixels) was 
achieved for a resolution of R $\approx$ 2200 over the spectral range 
from 3960 to 4800 \AA.

The B\&C spectrograph uses a long slit (4$'$).  All observations were
made using a slit width of 2.5$''$.  An average bias as well as sky emission
lines were removed by subtracting a median averaged image of several
unique slit positions.  Pixel-to-pixel gain variations on the CCD detector 
were removed using observations of an illuminated reflective spot 
inside the dome.  Observations of a Helium-Argon lamp taken 
periodically through the night provided the wavelength calibrations.
Integration times ranged from 16 minutes (Cyg OB2 A46) to as long as
an hour (Cyg OB2 A20).  The signal-to-noise ratio in the line
free continuum exceeds 50 for nearly all spectra.  A few spectra
drop just below this value at the shortest wavelengths where the 
CCD response is waning.

\begin{deluxetable}{lcccccc}
\tablewidth{0pt}
\tablecaption{Cygnus OB2 New MK Classification Spectra}
\tablehead{
\colhead{Star} &  
\colhead{$\alpha$(2000)} &
\colhead{$\delta$(2000)} &
\colhead{m$_{B}$} &
\colhead{Exp Time} &
\colhead{S/N} &
\colhead{SpType} 
}
\startdata
A20  & 20 33 02.9  &  40 47 25  & $\sim$14.5 &  60 min. & 50     & O8 II((f)) \\
A23  & 20 30 39.7  &  41 08 48  & $\sim$14.0 &  24 min. & 80     & B0.7 Ib \\
A27  & 20 34 44.7  &  40 51 46  & $\sim$13.0 &  24 min. & 100    & B0 Ia \\
A29  & 20 34 56.0  &  40 38 18  & $\sim$14.0 &  24 min. & 80     & O9.7 Iab \\
A32  & 20 32 30.3  &  40 34 33  & $\sim$14.0 &  44 min. & 60     & O9.5 IV \\
A34  & 20 31 36.9  &  42 01 21  & $\sim$13.0 &  24 min. & 100    & B0.7 Ib \\
A36  & 20 34 58.7  &  41 36 17  & $\sim$13.0 &  30 min. & 80     & B0 Ib(n) sb2? \\
A37  & 20 36 04.5  &  40 56 13  & $\sim$14.5 &  30 min. & 50     & O5 V((f)) \\
A39  & 20 32 27.3  &  40 55 18  & $\sim$14.0 &  30 min. & 60     & B2 V \\
A41  & 20 31 08.3  &  42 02 42  & $\sim$13.0 &  30 min. & 100    & O9.7 II \\
A42  & 20 29 57.0  &  41 09 53  & $\sim$14.5 &  50 min. & 60     & B0 V \\
A43  & 20 32 38.5  &  41 25 13  &  12.03     &  24 min. & 100    & O8 V(n)\tablenotemark{a}  \\
A44  & 20 31 46.0  &  40 43 24  & $\sim$13.5 &  30 min. & 60     & B0.5 IV \\
A45  & 20 29 46.6  &  41 05 08  & $\sim$13.0 &  30 min. & 80     & B0.5 V(n) sb2? \\
A46  & 20 31 00.1  &  40 49 49  & $\sim$12.0 &  16 min. & 80     & O7 V((f)) \\
\enddata
\tablenotetext{a}{SpType from Hutchings (1981) B0; Massey \& Thompson (1991, MT91) O7.5 V.}
\end{deluxetable}

\subsection{Target Selection}

A list of the stars observed and their positions is given in Table 1.
Stars were selected from the list of candidate early-type members given in
CPR2002.  Interstellar extinction towards this sample is
exceedingly high in the blue-optical, ranging from A$_B$ = 6 to more than
10 magnitudes.  Estimated $B$ magnitudes were calculated from Table 1 in
CPR2002, assuming a standard extinction law (Rieke \& Lebofsky
1985).  Observations were made of all candidate early-type stars 
with estimated B magnitudes brighter than $B \approx 14.5$.  No stars 
were observed from the list of Br$\gamma$ emission stars in CPR2002
(their Table 2).  

Optical spectra of the quality presented here were not previously 
available for any of the target stars with the exception of one 
star, Cyg OB2~A43.  This star was among the first OB stars found in 
Cyg OB2 based on photometric colors and was identified by Schulte 
as star 16 (Schulte 1956).  Hutchings (1981) and MT91 obtained 
spectroscopic measures of Cyg OB2~A43, 
assigning MK spectral types of B0 and O7.5 V, respectively.

\begin{figure}
\epsscale{0.8}
\plotone{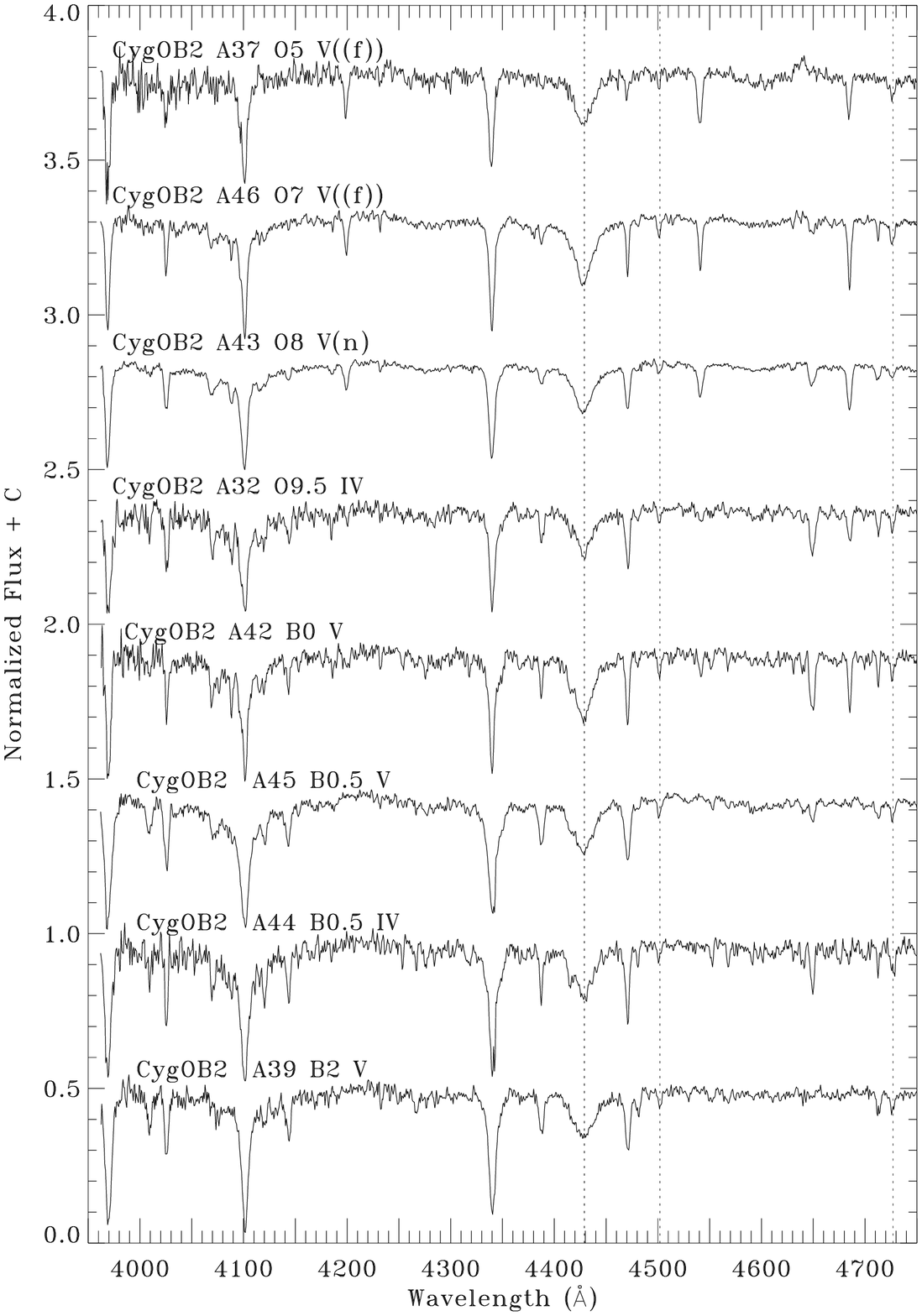}
\caption{MK Classification spectra of newly confirmed dwarf OB stars toward Cyg OB2.
Interstellar features due to diffuse interstellar bands are marked with a vertical
dashed line.\label{fig1}}
\end{figure}

\begin{figure}
\epsscale{0.8}
\plotone{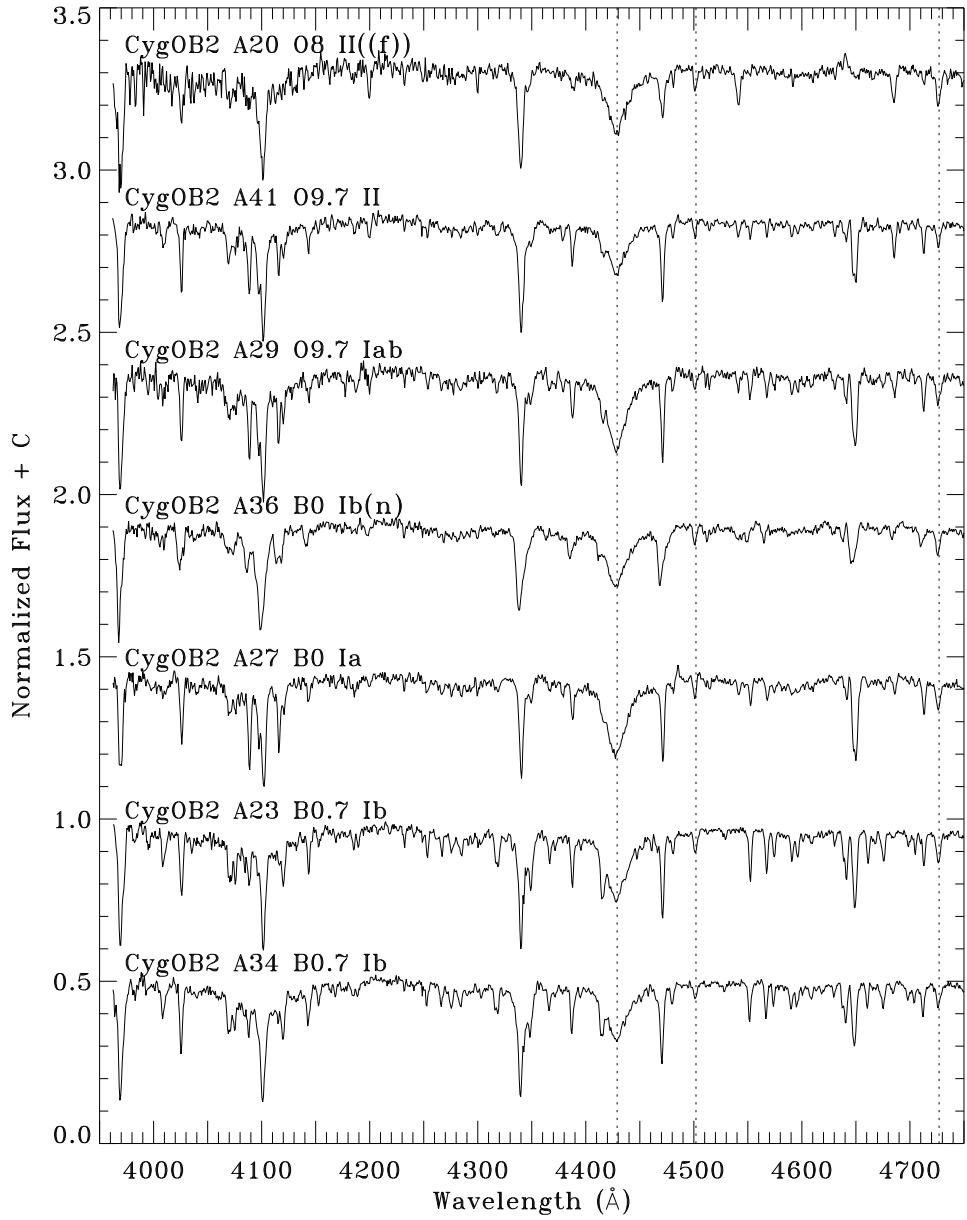} 
\caption{OB supergiant stars toward Cyg OB2, see Fig.\ 1.\label{fig2}}
\end{figure}

\section{THE SPECTRA}

The new optical spectra are shown in Figures 1 \& 2.  A very quick examination
shows that all stars from the sample show HeI 4471 \AA\, indicating
they are early-type stars, as predicted by CPR2002.  
MK spectral types, determined based on visual comparisons with 
the CCD spectral atlas of Walborn \& Fitzpatrick (1990), are given 
in Table 1.  There is no quantitative method for estimating the
uncertainty in spectral types, however, all 15 stars were independently
typed by three persons, myself, Phil Massey and Nolan Walborn. There
was excellent agreement within a spectral class for all stars (except
A39 and A44, which I typed differently).  The adopted classifications, 
independently and consistently given by both Massey and Walborn, 
are listed in Table 1.

Because of the large line-of-sight extinction, very strong interstellar
features are seen in our spectra.   The commonly detected diffuse interstellar
band (DIB) at 4428 \AA\ is shown to be exceedingly strong in all the
spectra.  A composite DIB spectrum, which was created from all the
Cyg OB2 star spectra after prominent stellar photospheric features
where removed, is shown in Figure 3.  Using the line lists provided by
Jenniskens \& Desert (1994), additional DIB features are clearly 
detected at 4501.80, 4726.59/4727.06 blend, 4761.67/4762.57 blend (one
narrow and one quite wide) and 4780.09 \AA.  Jenniskens \&
Desert (1994) also list a probable, very broad DIB centered at
4595.0 \AA.  We may be detecting such a feature, however the line
does lie in a region where weak Si III stellar features are plentiful
and  might not have been adequately removed in the composite spectrum
of Fig.\ 3.   Line centers, equivalent
width and FWHM measures have been made from fitting the spectrum in
Fig.\ 3. 
The fit is shown over-plotted in Fig.\ 3 and the results from the fit
are given in Table 2.

\begin{figure}
\epsscale{0.8}
\plotone{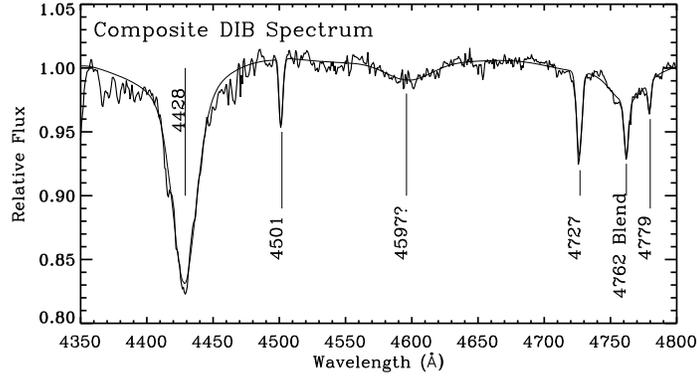}
\caption{An average spectrum of fifteen Cyg OB2 stars with prominent stellar
features removed to show the main diffuse interstellar band (DIB) features. 
A fit of the DIB features used to create inputs to Table 2 is shown over-plotted 
with the spectra.\label{fig3}}
\end{figure}

\begin{deluxetable}{lcc}
\tablewidth{0pt}
\tablecaption{Composite DIB Features}
\tablehead{
\colhead{Central} &  
\colhead{EQW\tablenotemark{a}} &
\colhead{FWHM\tablenotemark{b}}  \\
\colhead{$\lambda$, \AA} &
\colhead{\AA} &
\colhead{\AA} 
}
\startdata
4428.2  &  4.8   &  18  \\
4501.2  &  0.2   &  3.4 \\
4597.5  &  0.6   &  40  \\
4726.5  &  0.3   &  4.7  \\
4762.0  &  0.2   &  4.3  \\
4762.3  &  1.0   &  35  \\
4779.4  &  0.07  &  2.8 \\
\enddata
\tablenotetext{a}{EQW: Equivalent width of feature, measured in Angstroms.}
\tablenotetext{b}{FWHM: Full
width, half-maximum line depth, measured in Angstroms.}
\end{deluxetable}

\section{THE NEW OB STARS: ARE THEY MEMBERS OF CYG OB2?}

Kn\"{o}dlseder (2000) makes a convincing argument, based on 2MASS stellar
counts, that a significant fraction of the Cyg OB2 star cluster has been
missed in optical surveys.  Follow up work by CPR2002 appears to support
this interpretation.  The amazing success of the CPR2002 study to have
identified OB stars in the field would suggest that all 46 candidate OB
stars found are indeed OB stars.   However, a second equally important goal 
of this paper is to determine if the OB stars uncovered by CPR2002 are 
members of the optical OB cluster Cyg OB2 studied 
by MT91.  In this section we will compare the characteristics of the newly
identified OB stars with the characteristics of the previously identified 
OB stars of Cyg OB2 to determine this.

\subsection{Location of the new stars}

In Figure 4 is shown the CPR2003 survey area, 
covering a 1.5$^{\circ}$ x 2$^{\circ}$ field of view.  The center of the field 
is aligned with the center of the star distribution seen for 
Cyg OB2 by Kn\"{o}delseder (2000) using 2MASS, $\alpha = 20^{h} 33^{m} 10^{s}, \ 
\delta=
+41^{\circ} 15.7'$.  The circles correspond 
to the 120 stars previously identified to be members of the Cyg OB2 
cluster by MT91 based on UBV colors.  Asterisks in Figure 4 locate 
a total of 62 new massive star candidates identified by CPR2002.  
The asterisk sources include 45 candidate early-type members 
(CPR 2002 A43 is left out since it was previously identified as 
Schulte 16), 14 stars with Br~$\gamma$ emission (6 of 
their 20 Br~$\gamma$ sources were previously identified), and 3 new
sources showing CO in emission.

What is first immediately obvious in this image is that most of
the new sources found by CPR2002 lie well away from the 
previously identified optical cluster stars.  Only near the south 
and west side of the optical field are new candidate OB members seen to 
lie in close proximity to previously identified cluster members.  This study 
has confirmed 14 new OB stars in the field, all 14 of which are labeled with 
CPR2002 identifications in Fig\ 4 (A20, A23, etc.).  Of the 14 new OB stars 
confirmed in this study, all lie a significant distance from the previously 
identified optical cluster of MT91. The reason most of these OB stars were 
missed by MT91 was simply because the field over which MT91 searched did not 
extend to these angular distances.

\begin{figure}
\epsscale{0.8}
\plotone{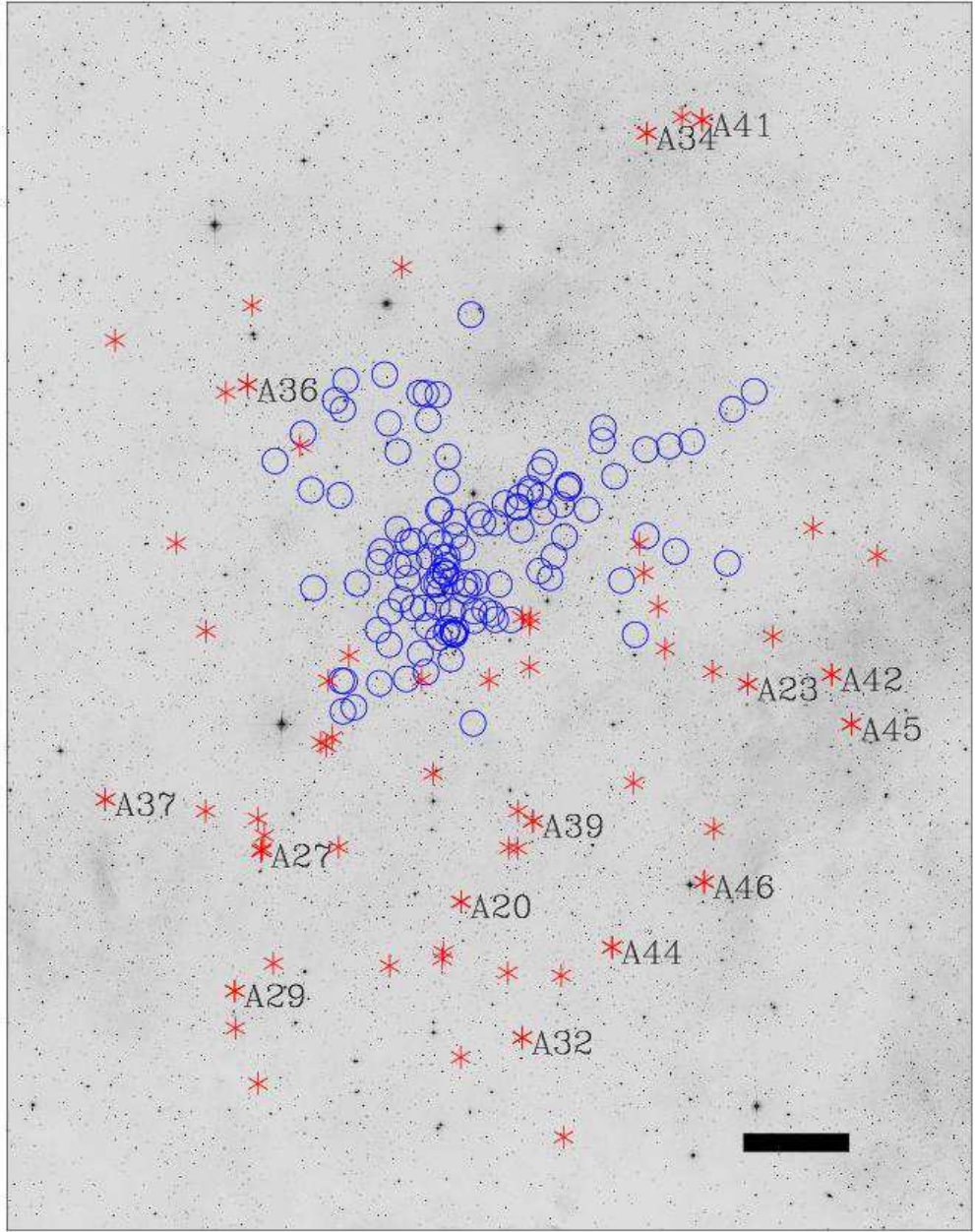}
\caption{A 1.5$^{\circ}$ by 2$^{\circ}$ optical field from
the Digitized Sky Survey centered on Cyg 
OB2 as defined by Kn\"{o}delseder (2000) using the 2MASS survey, (J2000) 
20$^h$33$^m$10$^s$ $+$41$^{\circ}$15.7$'$.  OB stars originally found 
by Massey \& Thompson (1991, MT91) are shown as circles. Asterisks indicate 
the location of new OB candidates recently uncovered by CPR2002. 
Stars from the CPR2002 survey which are confirmed in this 
study to be OB stars are marked with their star identifications as
listed in Table 1.  The legend bar in the lower right measures 10$'$.
\label{fig4}}
\end{figure}

\begin{deluxetable}{llccccr}
\tablewidth{0pt}
\tablecaption{The Dwarf Sample}
\tablehead{
\colhead{Star} &  
\colhead{SpTy} &
\colhead{T$_{eff}$} &
\colhead{Av} &
\colhead{m$_V$} & 
\colhead{M$_V$} & 
\colhead{V-M$_V$}
}
\startdata
059 & O8.5V & 37239 &  5.21 & 11.18 & -4.42 & 10.38 \\
070 & O9V & 35974 &  6.96 & 12.99 & -4.19 & 10.22 \\
145 & O9.5V & 34673 &  4.23 & 11.62 & -3.98 & 11.37 \\
187 & B0.5V & 28183 &  5.40 & 13.24 & -2.69 & 10.53 \\
215 & B1V & 25409 &  4.03 & 12.97 & -2.15 & 11.10 \\
227 & O9V & 35974 &  4.73 & 11.47 & -4.19 & 10.93 \\
250 & B1V & 25409 &  4.08 & 12.88 & -2.15 & 10.96 \\
258 & O8V & 38459 &  4.49 & 11.10 & -4.74 & 11.35 \\
259 & B0.5V & 28183 &  3.83 & 11.42 & -2.69 & 10.27 \\
292 & B1V & 25409 &  5.40 & 13.08 & -2.15 &  9.83 \\
299 & O8V & 38459 &  4.61 & 10.84 & -4.74 & 10.97 \\
317 & O8V & 38459 &  4.73 & 10.66 & -4.74 & 10.67 \\
339 & O8.5V & 37239 &  4.94 & 11.60 & -4.42 & 11.08 \\
376 & O8V & 38459 &  4.96 & 11.91 & -4.74 & 11.69 \\
378 & B0V & 31622 &  7.14 & 13.49 & -3.38 &  9.73 \\
390 & O8V & 38459 &  6.72 & 12.95 & -4.74 & 10.97 \\
421 & O9.5V & 34673 &  6.83 & 12.86 & -3.98 & 10.00 \\
429 & B0V & 31622 &  5.64 & 12.98 & -3.38 & 10.71 \\
455 & O8V & 38459 &  6.38 & 12.92 & -4.74 & 11.27 \\
467 & B1V & 25409 &  5.04 & 13.43 & -2.15 & 10.55 \\
470 & O9.5V & 34673 &  5.19 & 12.50 & -3.98 & 11.29 \\
473 & O8.5V & 37239 &  5.29 & 12.02 & -4.42 & 11.14 \\
480 & O7.5V & 39810 &  5.82 & 11.88 & -5.06 & 11.12 \\
485 & O8V & 38459 &  5.58 & 12.06 & -4.74 & 11.21 \\
507 & O8.5V & 37239 &  5.61 & 12.70 & -4.42 & 11.51 \\
515 & B1V & 25409 &  6.97 & 14.66 & -2.15 &  9.85 \\
531 & O8.5V & 37239 &  5.57 & 11.58 & -4.41 & 10.42 \\
534 & O7.5V & 39810 &  6.54 & 13.00 & -5.06 & 11.51 \\
555 & O8V & 38459 &  6.57 & 12.51 & -4.74 & 10.68 \\
588 & B0V & 31622 &  6.02 & 12.40 & -3.38 &  9.76 \\
605 & B0.5V & 28183 &  4.23 & 11.78 & -2.69 & 10.24 \\
692 & B0V & 31622 &  5.69 & 13.61 & -3.38 & 11.30 \\
696 & O9.5V & 34673 &  5.85 & 12.32 & -3.98 & 10.45 \\
716 & O9V & 35974 &  6.10 & 13.50 & -4.19 & 11.59 \\
736 & O9V & 35974 &  5.46 & 12.79 & -4.19 & 11.52 \\
 &  &  & \multicolumn{4}{r}{Average Vo - Mv =} 10.80 \\
\enddata
\end{deluxetable}

\subsection{Extinction towards Cyg OB2}

From the full sample of 71 Cyg OB2 stars with spectral types published in MT91, 
a subset of 35 stars, listed in Table 3, was selected to perform specific 
calibrations of the cluster characteristics. This subset of stars was selected 
to have optical spectral types between O7.5 and B1 inclusive, all are 
normal dwarf stars, and all have both UBV values as well as 2MASS JHK$_S$ 
colors.  None of the new OB stars presented in this paper were included.
Using the intrinsic UBV color terms given by FitzGerald (1970; and recently
used by Slesnick, Hillenbrand, \& Massey  2002) 
and JHK$_S$ color terms given by Lejenue \& Schaerer 
(2001) for dwarf stars,  the extinction characteristics were determined
toward each star independently based on the stars observed UBVJHK$_S$ 
colors.  This has lead to a well constrained average extinction law 
towards the cluster as illustrated in Figures 5, 6 and 7.  The extinction
law created from the fits shown
in Figs.\ 5, 6, and 7 are given in Table 4.   The optical
extinction law was previously measured by MT91 and Torres-Dodgen et al.\ (1991).
As seen in these previous studies, the extinction was measured to
be consistent with a total to selective extinction ratio, $R_V$, of 3.0.  
The extinction law through the near-infrared matches that given for the 
general interstellar medium by Rieke \& Lebofsky (1985), except for the K band.  
The centroid of the 2MASS K$_S$ band is at 2.16 $\mu$m, while Rieke \& Lebofsky used 
Johnson K-band filters centered at 2.22 $\mu$m.  This leads to a slightly larger 
extinction ratio in the 2MASS K$_S$ band ($A_{K_s}/A_V = 0.125$) relative to the 
Rieke \& Lebofsky (1985) K band ($A_K/A_V = 0.112$).

\begin{deluxetable}{ccc}
\tablewidth{0pt}
\tablecaption{Insterstellar Extinction Law toward Cyg OB2}
\tablehead{
\colhead{$\lambda$} &  
\colhead{$A_{\lambda}/A_V$} &
\colhead{RL85} 
}
\startdata
U (0.36 $\mu$m) &  1.600   &  1.531  \\
B (0.44 $\mu$m) &  1.333   &  1.324  \\
V (0.55 $\mu$m) &  1.000   &  1.000  \\
J (1.24 $\mu$m) &  0.282   &  0.282  \\
H (1.66 $\mu$m) &  0.175   &  0.175  \\
K (2.16 $\mu$m) &  0.125$^1$   &  0.112$^2$  \\  
\enddata
\tablenotetext{1}{K-band centroid at 2.16 $\mu$m.}
\tablenotetext{1}{K-band centroid at 2.22 $\mu$m.}
\end{deluxetable}

\begin{figure}
\epsscale{0.8}
\plotone{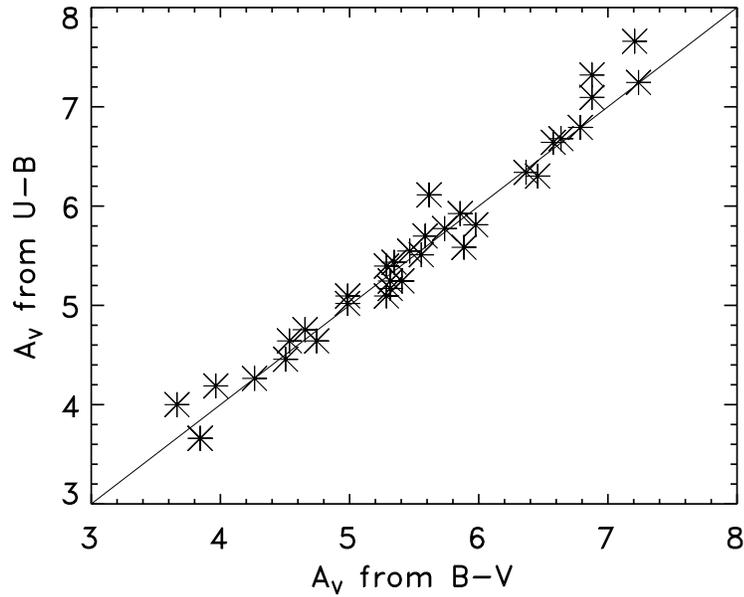}
\caption{UBV color excesses for 35 OB dwarf stars in Cyg OB2.  See \S4.2 \label{fig5}}
\end{figure}

\begin{figure}
\epsscale{0.8}
\plotone{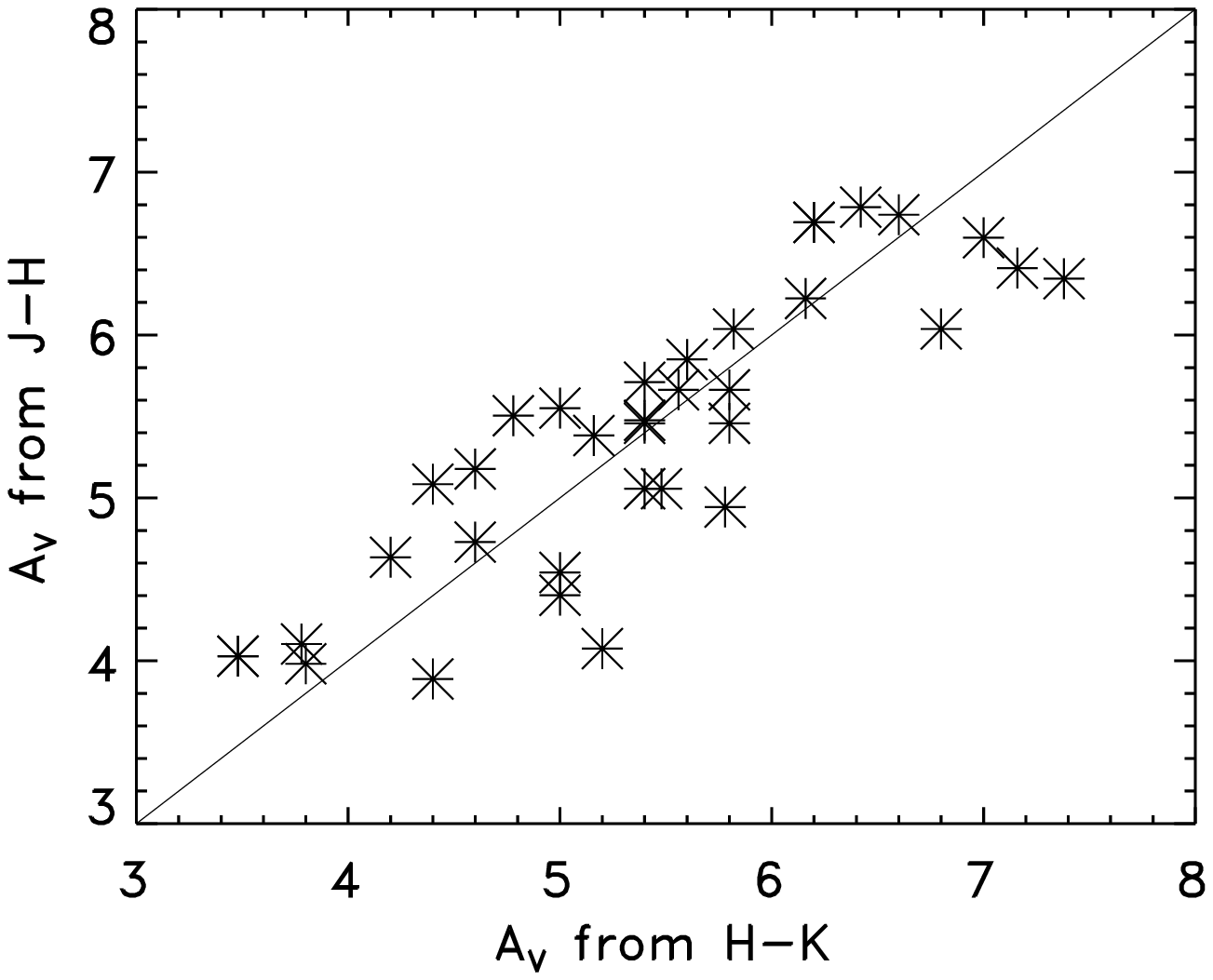}
\caption{JHK$_S$ color excesses for 35 OB dwarf stars in Cyg OB2. \label{fig6}}
\end{figure}

\begin{figure}
\epsscale{0.8}
\plotone{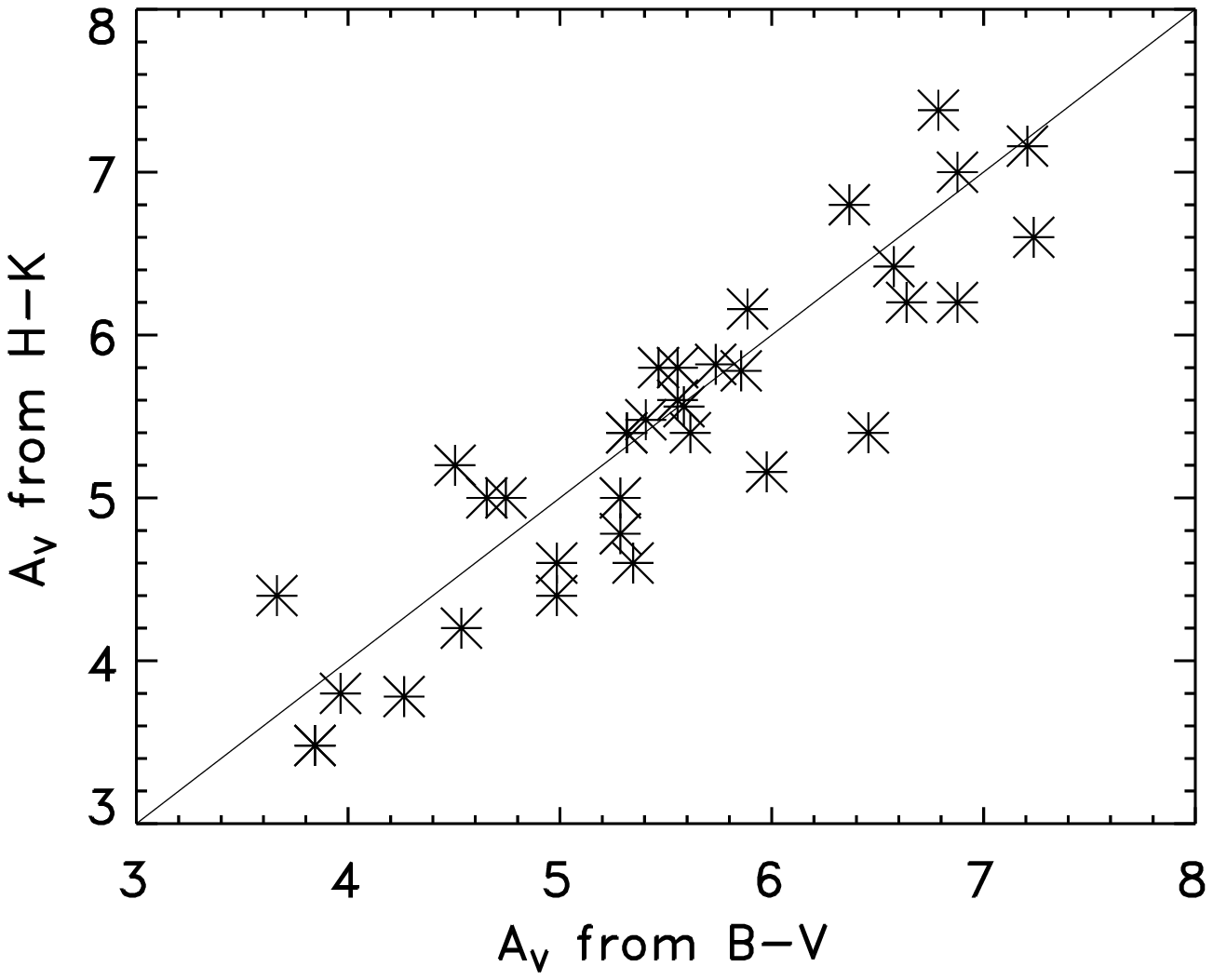}
\caption{BVHK$_S$ color excesses for 35 OB dwarf stars in Cyg OB2. \label{fig7}}
\end{figure}

With a well established extinction law derived for the cluster, A$_V$ values for every
star with optical spectral types in our full sample of 85 (71 previously identified by 
MT91, plus the 14 new OB stars confirmed in this study) could be calculated from either 
UBV, JHK$_S$ colors, or both depending on what was available.  Extinction values for all 
stars in this study (inclusive of the 35 dwarf stars also listed in Table 4) are given 
in Table 5.  Of the 
85 stars in our complete sample extinction varied from 
A$_V$ = 3.83 up to A$_V$ = 10.70 for the infamous B5 supergiant, IV Cyg \#12.

It is worth noting that 
the average extinction towards the 14 CPR2002 OB stars was not significantly greater
than that found towards the OB stars previously identified in the optical by MT91
Also, the new OB stars did not show an extinction law any different from that 
determined for the general Cyg OB2 cluster, beyond a few intrinsically unusual 
stars listed below.  The former indicates that high extinction 
was not the reason these 14 stars were missed in the optical surveys.  However, the
remaining 31 stars in the CPR2002 sample of candidate OB stars {\sl do} have significantly
higher extinction than those observed for this study.  This may have contributed 
to their being missed in the optical study of MT91.

\subsubsection{Sources with unusual photometric properties}

Once the extinction law had been derived for the sight line, it was easy to 
spot outliers in the numerous color-color diagrams created for the entire 
sample. Listed below are six sources for which their photometry showed 
unusual values. With the exception of these six stars, the extinction law towards
all the stars in the sample, those found by MT91 and those uncovered by
the CRP2002 survey, have essentially identical extinction properties.  

{\bf MT91 390.}  This star was identified by Pigulski \& Kolaczkowki (1998) 
as having an irregular variability in the I-band, but otherwise looks fine 
in our colors. This star was classified as an O8V star by MT91.

{\bf MT91 556} Schulte 18. This star was classified as B1Ib by MT91, however it 
shows some peculiarities in its UBV colors.  Schulte 18 had been identified by 
Pigulski \& Kolaczkowki (1998) as having an irregular and rather large variability 
in the I-band. Unfortunately, the spectrum obtained by MT91 is not shown in 
their paper.

{\bf MT91 575} CPR2003 B13.  This star was previously known to be an emission 
line star from Merrill \& Burwell (1950).   Its spectral type from MT91 is 
B1.5V.  This star has the largest IR excess in the entire sample, as seen in
our (H-K$_S$) vs (J-H) plots.  The star is listed in HAe/Be star catalog 
of The et al.\ (1994).  The spectrum as shown in MT91 looks normal.

{\bf MT91 605.} This star was previously identified to be an emission line star 
by Merrill \& Burwell (1950).  The UBV and JHK$_S$ colors for this star appear 
normal.  The spectral type, B0.5V, comes from MT91.  The spectrum of this star, 
as shown in MT91 also appears normal.  

{\bf MT91 793} CPR2002 B16. This star shows a strong near-infrared excess in 
its JHK$_S$ colors.  The spectral type from MT91 is uncertain (B1.5III?).  The 
spectrum 
shown in MT91 has numerous emission lines, both in hydrogen and from other 
transitions.  It seems likely that this is a Be type star, though it is
not mentioned as such in the literature.

{\bf CPR2003 A34.} This star shows a relatively strong near-infrared excess 
in the (H-K$_S$) vs (J-H) diagram. From this work, the star has been classified 
as a B0.7 Ib star.

\subsection{Distance to Cyg OB2}

The subsample of dwarf late-O and early-B stars given in Table 3 was used to 
derive the cluster distance.  Calculating a star's distance requires knowing
its apparent magnitude, line of sight extinction and its absolute magnitude.  
The first two measures are well constrained for our sample.  The greatest uncertainty 
in obtaining the distance will be in choosing the absolute magnitude for the
dwarf star sample.

There are numerous tables of absolute magnitude as a function of spectral type 
available in the literature.  These show relatively good agreement over the past 3
decades (Walborn 1973; Humphreys \& McElroy 1984; Conti 1988; Vacca et al.\ 1996).  
Because of the large number of dwarf 
stars with similar age and distance, the Cyg OB2 data set can be used to 
test the consistency of the absolute magnitude calibration.
First, the distance modulus was determined using the absolute magnitudes of
Massey and collaborators, most recently published in Slesnick et al.\ 2002,
based on earlier studies by Conti (1988) for the O stars and 
Humphreys \& McElroy (1984) for the B stars.   
These values for absolute magnitude are comparable
to what has been typically used in the literature, dating back
as far as Walborn (1973) and are within a tenth of a magnitude of
those quoted in Vacca et al.\ (1996). 
Figure 8 shows the distance modulus of each
star from the subsample (35 dwarf stars) as a function of spectral type 
(converted to T$_{eff}$).  The average of all 35 stars yields a distance 
modulus of 11.16.  This is very similar to what was 
determined in MT91 (DM = 11.20).   Some structure is seen
in Fig.\ 8, however.  While the O stars (Log T$_{eff} > 4.50$) give 
a consistent distance, a slight increase is seen in the predicted 
distance modulus with the cooler stars.  If just the O stars are 
averaged, a slightly lower value of DM = 10.99 is obtained.

\begin{figure}
\epsscale{0.7}
\plotone{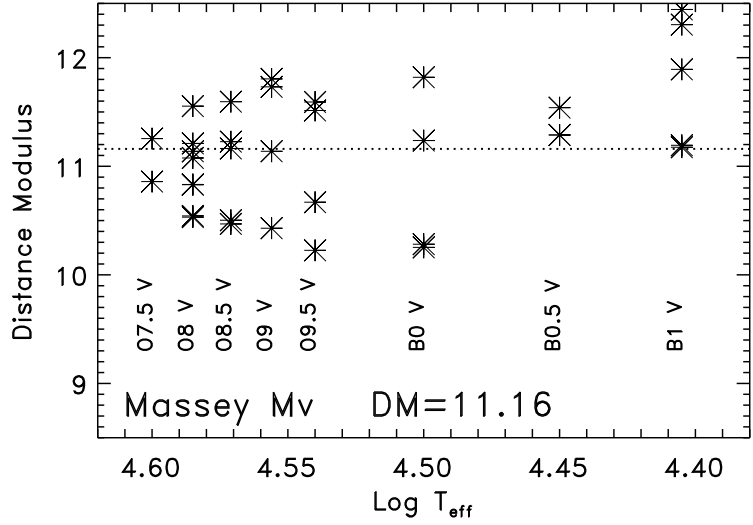}
\caption{The distance modulus calculated using the spectral type, effective temperature
and absolute magnitude relations presented in Slesnick, Hillenbrand \& Massey 
(2002).\label{fig8}}
\end{figure}

The distance to Cyg~OB2 was also calculated using the Hipparcos derived 
absolute magnitudes given by Wegner (2000).  The average value for the
distance modulus, DM = 10.07, and shown in Fig.\ 9, is considerably 
closer than originally given by MT91.  Could the distance modulus 
derived from the Hipparcos values be correct?  Comparisons to nearly every 
other study of absolute magnitude as a function of spectral type in 
OB stars in the last few decades shows the Wegner values for absolute
magnitude to be 
significantly lower, typically by a full magnitude in mid and
late O and early B stars.  This strongly suggests that the Hipparcos 
distances have a strong ``near'' bias not fully appreciated.  Fig.\ 9 shows
a weak but significant dependency of spectral type with average distance, 
in the same sense as is seen in Fig.\ 8.  Such a dependency would be introduced if 
the age of the stars used in the calibration is not taken into consideration.  
Because Cyg OB2 is only a few million years old, the cooler dwarf stars 
would be found exceedingly close to the ZAMS.  Typical late-O and 
early-B dwarfs used for the calibration of $M_V$ exist in clusters 
less massive and typically older than Cyg OB2.  While these older dwarf
stars may have the same spectral type and luminosity class, they represent 
slightly more massive stars with higher intrinsic luminosities than our 
near-ZAMS B stars.  This effect, also discussed by Walborn (2002), is 
borne out in the increased DM measured as a function of spectral type in 
both the Slesnick et al.\ (2002)
absolute magnitudes (Fig.\ 8) and the Hipparcos values (Fig.\ 9).

\begin{figure}
\epsscale{0.7}
\plotone{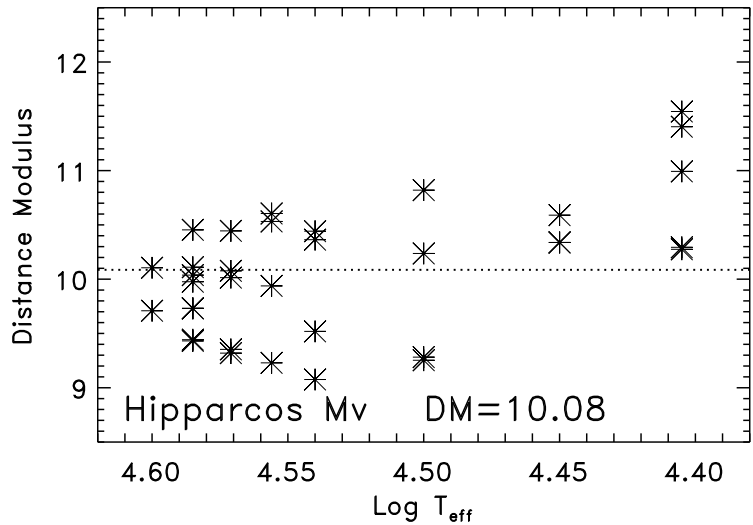}
\caption{The distance modulus calculated from every star listed in Table 4 using
absolute magnitudes determined from Hipparcos (Wegner 2000) with effective
temperature scales of Slesnick et al.\ (2002).
\label{fig9}}
\end{figure}

\subsubsection{Effective Temperatures of OB stars}

Fig.\ 8 and 9 demonstrate that some consideration of the very young age of 
the Cyg OB2 cluster needs to be considered in order to properly determine 
its distance modulus.  If Cyg OB2 is to be fit to a very young isochrone, 
there are just two things that can change in Fig.\ 8 and 9 to make a
better fit.  Either 
the absolute magnitudes are correct and the temperature scales needs
to be changed or 
the other way around.  This then introduces a second calibration which is 
equally important to estimating distance modulus: the stellar effective
temperature scale.
Knowing a star's effective temperature is crucial in estimating its absolute 
luminosity, particularly in theoretical models.  Perhaps the most widely used 
reference to this relationship is given by Vacca et al.\ (1996) 
based on calculations using plane parallel, pure hydrogen and helium,
non-LTE atmosphere models.  However, several 
independent groups have recently shown that more sophisticated atmospheric models lead
to significantly lower effective temperatures in the O-type stars, both in 
the dwarf stars (Martins et al.\ 2002; Markova et al.\ 2003) and the
giant and supergiant O supergiants (Bianchi \& Garcia 2002; Herrero, Puls 
\& Najarro 2002).   The shift in the spectral-type effective-temperature scale 
is most pronounced in mid-O dwarf stars, and diminishes to near zero offset 
from the Vacca et al.\ estimates of effective 
temperatures for early-B stars.  The implications of a new, lower effective 
temperature scale, as discussed in Bianchi \& Garcia (2002), includes a 
lower luminosity, thus lower mass, when compared to evolutionary tracks.
Since the typical goal in studying most OB clusters is to compare the stars 
to theoretical evolutionary tracks, a lowering of the O star temperature scale 
will have an immediate effect on results for age, mass and distance to OB
clusters.

\begin{figure}
\epsscale{0.7}
\plotone{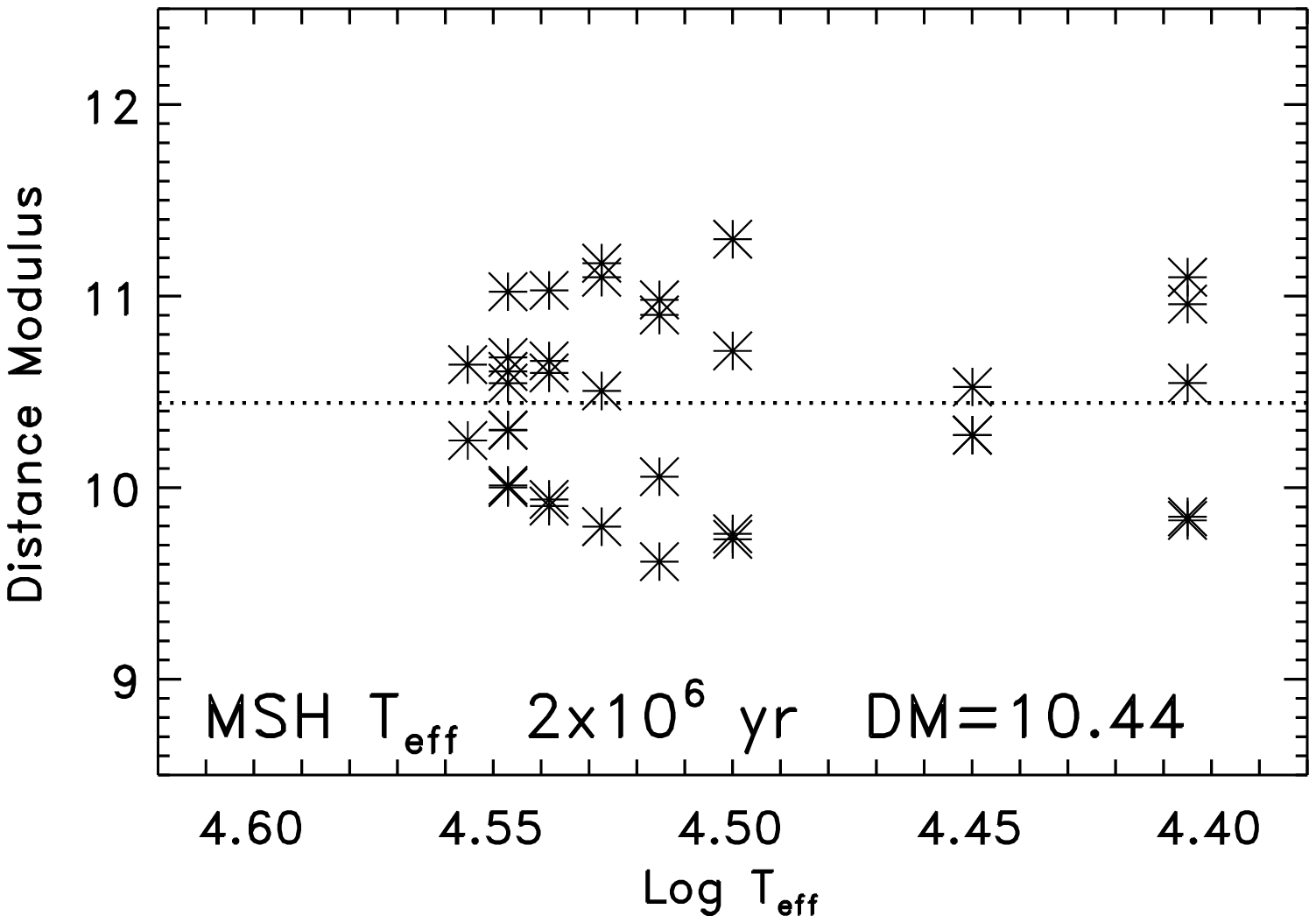}
\epsscale{0.8}
\caption{The distance modulus calculated using the effective temperature
relations of Martins, Schearer and Hillier (2002) with absolute magnitudes
scaled to match a 2 million year isochrone. 
\label{fig10}}
\end{figure}

\begin{figure}
\epsscale{0.7}
\plotone{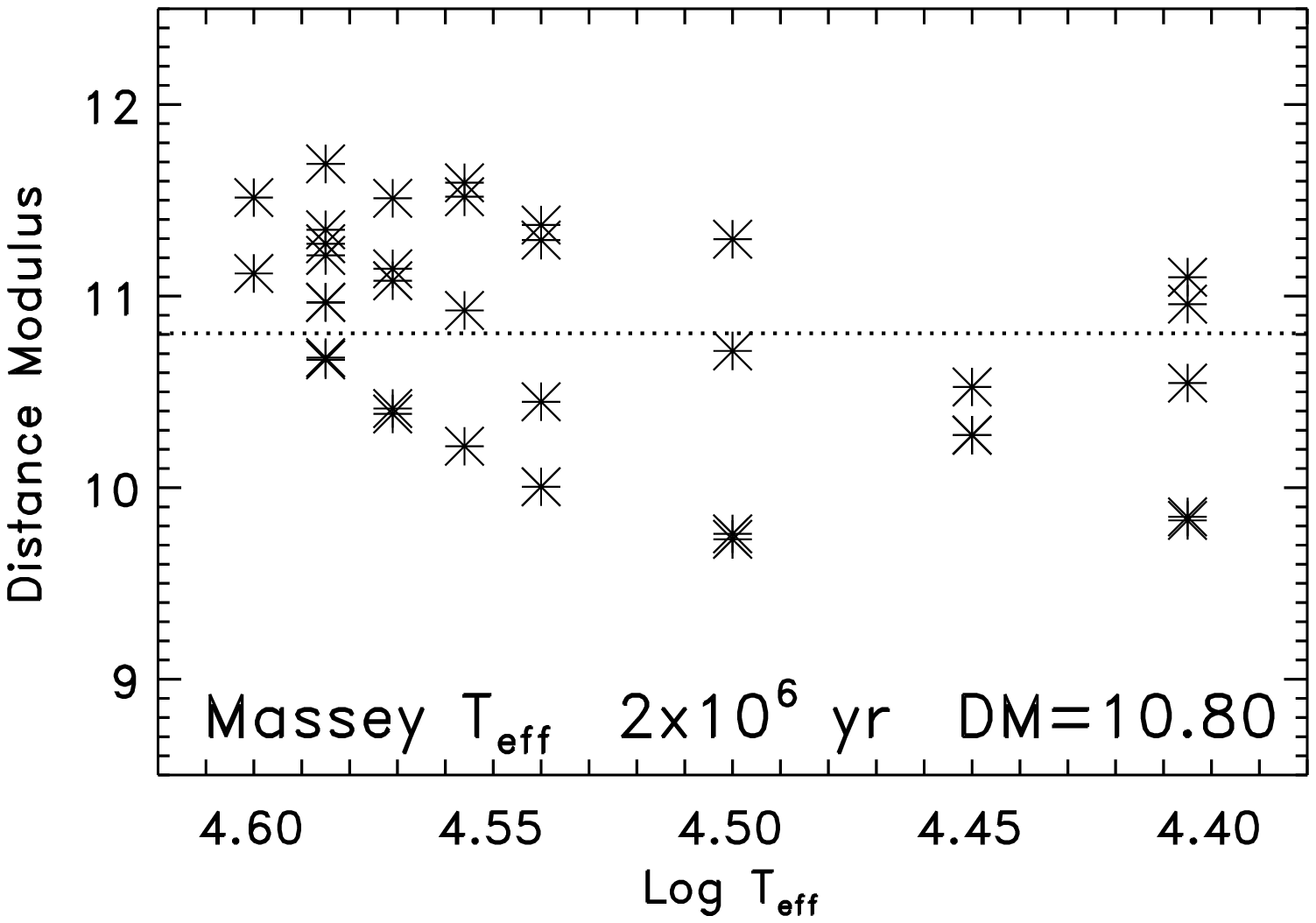}
\caption{The distance modulus calculated from every star listed in Table 4 using
the effective temperature scale of Slesnick et al.\ (2002) with the absolute magnitudes
forced to match a 2 million year isochrone.
\label{fig11}}
\end{figure}

It is worth noting,
the reduced stellar effective temperatures as proposed by recent
theoretical arguments appear drastic only when these new values are compared to 
Vacca et al.\ (1996).  As discussed by Massey \& Hunter (1998) and Massey et 
al. (2000), the Vacca et al.\ values 
represented a significant upward trend from previous effective temperature 
scales (Conti 1973; Humphreys \& McElroy 1984; Chlebowski \& Garmany 1991).
In fact, the Vacca et al.\ absolute magnitudes, bolometric
corrections, and many other calibrations are all highly generous with respect
to previous estimates of all these values. Recent studies of young open clusters 
in the Large Magellanic Cloud (LMC) confirm this.  Heydari-Malayeri et al.\ (2003) showed 
a marked discrepancy between the high luminosities suggested by Vacca et al. 
(1996) for their cluster stars over those they derived 
directly from photometry and a known LMC distance.   A comparison of the
effective temperature and spectral type calibrations from the previous
decade (most notably Humphreys \& McElroy 1984) give good
agreement, though not entirely coincident, with the dwarf model atmosphere 
temperatures proposed by Martins et al.\ (2002).

Our ultimate goal in studying this cluster is to place the stars 
onto the HR diagram to determine distance, age, and stellar masses.
It is reasonable to assume the best fit would be 
one where the 35 dwarf stars line up along a single isochrone 
on the HR diagram, provided the stars were predominately co-eval
in formation.
Using the new spectral type T$_{eff}$ 
calibration given by Martins et al., we employed the Z = 0.02,
2x mass loss isochrones of LeJeune \& Schaerer (2001) to predict
total luminosity as a function of temperature for the dwarf stars.  
An isochrone of 2 million years was chosen, though the cluster clearly
shows an age spread of from 1 to 3 million years (see \S 4.4.1).  
Using the bolometric 
corrections of Slesnick et al.\ (2002), this gave a set of absolute V 
magnitudes as a function of effective temperature.  For the early B-stars, 
a seamless extrapolation from the Martins et al. values was created 
based on T$_{eff}$ scales given by Humphreys \& McElroy
(1984) and Chlewbowski \& Garmany (1991).  A third plot, showing the distance 
modulus as a function of T$_{eff}$ using the absolute magnitudes
determined in this way is shown in Figure 10.  The temperature dependence 
seen in Figs.\ 8 and 9 is now entirely gone.  Not surprisingly,
the average distance, DM = 10.44, is less than previous measures.  
The reduced stellar temperatures 
leads to a reduced absolute luminosity for O dwarf stars.  Distances 
predicted using O dwarf stars with such low absolute magnitudes can be
tested against other distance measures, such as 
to the Large Magellanic Cloud. Here, OB stars are routinely observed 
(Walborn \& Blades 1997; Massey, Waterhouse \& DeGioia-Eastwood 2000,
Heydari-Malayeri et al.\ 2003) and there are independent checks on 
distance to derive the absolute magnitudes.

\subsubsection{Absolute Magnitudes for ZAMS OB Stars}

If the temperature scale for OB stars as given in Figs.\ 8 and 9 is 
correct, we can improve the fit of Fig.\ 8 based on Slesnick et al.\
(2002) effective  
temperatures by making small adjustments to the absolute magnitude 
scale.  Again, the Z = 0.02, 2x mass loss isochrones of LeJeune 
\& Schaerer (2001) where used to provide a smooth predicted total 
luminosity as a function of temperature for the dwarf sample.  
An isochrone of 2 million years was chosen.  Using the bolometric 
corrections of Slesnick et al.\ (2002) with the original effective
temperature scale of MT91 (also listed in Slesnick, but these are
based primarily on Humphreys \& McElroy
1984, and Conti 1988), this provides a set of absolute 
V magnitudes.  This final set of spectral type, effective temperature
and absolute V magnitude, $M_V$, used for the dwarf sub-sample is 
listed in Table 3.  The DM value 
of each star is shown as a function of effective temperature 
(spectral type) in Fig.\ 11 and is also given in Table 3. The average
distance modulus determined in this way was found to be 10.80. 
Fig.\ 11 shows a weak trend not yet seen.  The hotter the star, 
the greater the predicted distance modulus, quite opposite to what 
was seen in Fig.\ 8 and 9.  One possible explanation is that the
cluster is older than 2 million years, and the hotter stars are
further from the ZAMS than we've estimated.  However, its clear based
on the presence of the extremely massive O3 If, and other early O
supergiants, that the cluster must indeed be quite young. Another explanation 
is that the assumed effective temperatures of the mid-O stars is too
hot, giving them too great a luminosity and thus placing them at
a further distance.  It will be possible to show in the next section
that explanation number one is ruled out.  Explanation number two 
lends a bit of support to the notion that traditional temperature 
scales for mid-O stars are a bit too high, even after coming down from
the generous temperature scale of Vacca et al.\ (1996).

\subsubsection{Final word on the Distance to Cyg OB2}

Using the traditional set of absolute magnitudes and stellar
effective temperatures for Cyg OB2 yields a distance modulus for the
cluster which is consistent with previous studies (MT91, Torres-Dodgen
et al.\ 1991).  However, weak but significant dependences of the 
distance calculated with the stellar spectral type are shown in many 
of our figures (Figs.\ 8, 9, and 11 in particular), indicating possible
room for improvement.  It is of interest to note that any attempt 
at better fitting the DM values as a function of spectral type 
ultimately leads to a closer distance for Cyg OB2.  It is perhaps 
premature to boldly accept the 
much closer distance for Cyg OB2 suggested by Fig.\ 10 using the 
Martins et al.\ (2002) derived values for effective temperature.
Beyond the possible inconsistencies with other measures of OB
distances as mentioned above,  there are significant ramifications 
to applying a new, closer distance to the Cyg OB2 cluster.  Perhaps 
the most important is a close distance reduces the luminosity of the 
early O supergiants in Cyg OB2.  Among the most important and best studied 
of the Cyg OB2 supergiants are the O3 If$^*$ star, IV Cyg \# 7 ([MT91] 457), 
and the O5 If stars, IV Cyg \#11, \#8A, and \#8C ([MT91] 734, 465, and 483, 
respectively).  In a recent paper, one in a series of papers studying 
the Cyg OB2 supergiant stars, Herrero et al.\ (2002) assume a distance 
of 1700 pc (the MT91 DM value of 11.2) for deriving the luminosity of 
these stars and in modeling their atmospheres and mass loss.  Moving 
the distance to just over 1200 pc (DM=10.44) reduces the intrinsic luminosity
of the O supergiant stars by a factor of two!  This low a luminosity is
simply not consistent with current wind theory because of the difficulty 
driving the mass loss seen in the Cyg OB2 supergiant stars if their 
luminosities were reduced a factor of two.  However, a
DM = 10.80 is not such a problem.  The derived luminosities of the 
before mentioned supergiants, O3 If$^*$ star, IV Cyg \# 7 ([MT91] 457), 
and the O5 If stars, IV Cyg \#11, \#8A, and \#8C, (MT[91] 734, 465, and 483, 
respectively) using a DM = 10.80 is given in Table 5.  For these stars,
we derived $Log \frac{L}{L_{\odot}}$ = 5.98, 6.06, 6.22, and 5.83,
respectively.  Herrero et al.\ (2002), derived $Log \frac{L}{L_{\odot}}$ = 
5.91, 5.92, 6.19, and 5.66 respectively.  Moreover, the luminosities 
calculated by Herrero et al.\ (2002) for stars [MT91] 632, 217, and
83 of $Log \frac{L}{L_{\odot}}$ = 5.77, 5.41, and 4.85, are a very good
match the values given in Table 5 for these stars: 5.80, 5.83, and
4.64, respectively.   For the remainder of this work, 
a distance modulus of 10.80 will be used for the Cyg OB2 cluster.

\subsection{The HR Diagram for Cyg OB2}

A new HRD for Cyg OB2 using the distance modulus of 10.80 is shown in Figure 12.
Here it can be seen how the location of the dwarf stars are well fit to a 2 million 
year old isochrone, as given by LeJeune \& Schaerer (2001), with a spread in 
age of between 1 and 3 million years old.  The luminosity 
spread seen in the B1.5  and B2 dwarfs looks to be greater and in the direction of
being over-luminous than is seen for the O and earlier B dwarfs.  There may be 
increased contamination from foreground stars at these dwarf star masses which
are slightly lower than their hotter dwarfs.

\begin{figure}
\epsscale{0.8}
\plotone{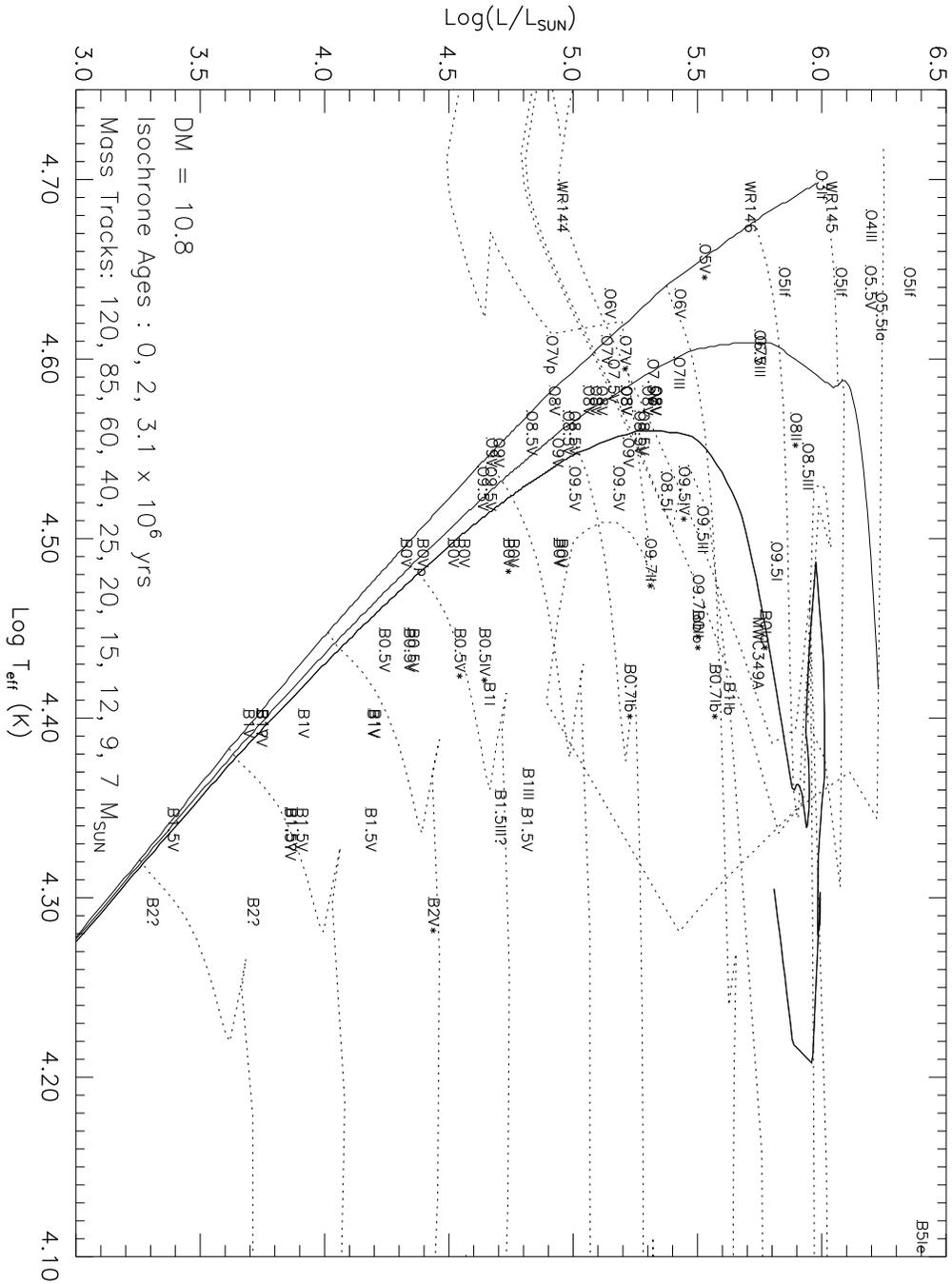}
\caption{A revised HR diagram for Cyg OB2 assuming the new distance of 1500 pc
(DM=10.8).  Newly identified stars from this study are identified with an asterisk
following their spectral types.  The distance of 1500 pc
was determined through main sequence fitting of the previously known (none of those
found in this study) O7.5 through B1 dwarfs to a 2 
million year old isochrone using effective temperatures given by Slesnick 
et al.\ (2002). See Fig.\ 11.
\label{fig12}}
\end{figure}

\subsubsection{Age of the Cyg OB2 Cluster}

As first stated in MT91, a reasonably well-defined main sequence is seen in the
HRD of Cyg OB2.  However, contamination is present in the form of several 
evolved giant and supergiant stars (Fig. 12).  This indicates contamination 
from a non-coeval population within the Cyg OB2 sample region.  Despite this
contamination, we can concentrate on the dwarf stars to estimate the approximate
age of the primary constituents of the cluster.  The presence the O5 and O5.5 
dwarf stars indicates an age not greater than 1 to 1.5 million years old.  
However, the O5 dwarf (A37) comes from the CPR2002 study and has yet to 
be confirmed 
as being part of the main Cyg OB2 cluster.  Furthermore, the luminosity of 
the O5.5 V ([MT91] 516), puts it far from the main sequence (or possibly at a 
nearer distance), making it also suspect as a reliable age measure for the bulk of
the cluster.  A well behaved population of dwarf stars is found starting 
at around O6 and O7 and by O7.5 and O8 there exists a particularly numerous 
sample of dwarf stars. These mid-O dwarfs put a strong upper limit on the age 
of the cluster of about 3 million years.  An age of 2 million years, with a spread 
of perhaps one million years, represents a reasonable age for the bulk of 
the Cyg OB2 cluster based on its most massive dwarf stars.  Indeed, a 
young age of not more than a few million years is needed to 
explain several very high luminosity blue supergiant stars (such as Cyg OB2 \#7,
the O3 If) and the possible Wolf-Rayet star members, the positions of which are 
also shown in Fig.\ 12.    The Wolf-Rayet stars where placed on 
the HRD in Fig.\ 12 based on observations of the stars apparent magnitude
and line of sight extinction as measured by
Schmutz \& Vacca (1991) for WR145; Dougherty 
et al.\ (2000) for WR146 and Massey et al.\ (2001) and references therein 
for WR~144. The temperature of the three WR stars were simply assumed to be 50,000 K.   
The temperature, apparent magnitude
and line of sight extinction of MWC349A was estimated by Hofmann et al.\ (2002).  
Citing proximity arguments, 
Massey et al.\ (2001) dismissed the Van der Hucht et al.\ (1981) claim 
that all three WR stars where members of Cyg OB2.  Only WR~144
is near the optically distinct cluster as first studied by MT91.  However,
if a new extended radius for the Cyg OB2 cluster is to be considered, then
WR~145, WR~146 and possibly MWC~349A should be re-considered as possible
members.

\subsubsection{Mixed-aged Stars towards Cyg OB2}

Let's consider the possibility of non-coeval members appearing in the HRD shown
in Fig.\ 12.
Over the temperature range Log T$_{eff}$ = 4.30 to 4.37, four early 
B stars lie well above the main sequence, A39 (B2 V), 642 (B1 III), 
575 (B1.5V), and 793 (B1.5III?) in Fig.\ 12.  Two show near infrared 
excesses (575, 793), consistent with them being Be stars.  A39 is likely
foreground.   An additional 11 giant and supergiant stars, lying mostly 
in the temperature range Log T$_{eff}$ = 4.40 to 4.53, are undeniably 
older than the few million years we've assigned to the dwarf stars.  
Of these eleven older, evolved stars, over half of them 
originated from the CPR2002 survey (A32, A23, A27, A29, A36, A41).  
The other five, S03, and MT[91] 632, 601, 642, and 83 where from the original 
optical survey.  The O5 V((f)) star identified by CPR2002 as A37, seems to be
two young for the remainder of the Cyg OB2 cluster, indicating an age 
below 1 million years.  Summing things up, 7 of the 14 new stars identified
by CPR2002 show ages inconsistent with the remainder of the cluster, and 
are unlikely members of Cyg OB2.  Extending the search 
for cluster members to larger distances from the central core has lead to 
an increase in contamination from non-member stars.  The IMF values and 
stellar counts made by MT91 still remain the best estimates for the cluster 
until the remaining OB candidates identified by CPR2002 are observed.  
However, it is clear that in an effort to move towards greater completeness 
of the cluster members, we have also moved towards a greater incidence of 
contamination.  Without reliable measures to establish membership, 
this will be a chronic problem for all studies attempting deep photometric 
studies of distant clusters within our galaxy.

\begin{deluxetable}{llccccc}
\tablewidth{0pt}
\tablecaption{Cyg OB2 Stars with Spectral Types}
\tablehead{
\colhead{Star} &  
\colhead{SpTy} &
\colhead{T$_{eff}$} &
\colhead{mv} &
\colhead{Av} & 
\colhead{BC} &
\colhead{Log$\frac{L}{L_{\odot}}$} 
}
\startdata
005 & O6V & 43551.2 & 12.93 &  6.19 & -3.73 &  5.13 \\
021 & B2? & 19952.6 & 13.74 &  4.41 & -2.00 &  3.29 \\
059 & O8.5V & 37239.2 & 11.18 &  5.21 & -3.38 &  5.26 \\
070 & O9V & 35975.0 & 12.99 &  6.96 & -3.31 &  5.20 \\
083 & B1I & 26302.7 & 10.64 &  4.03 & -2.67 &  4.64 \\
138 & O8.5I & 34435.0 & 12.26 &  6.73 & -3.25 &  5.36 \\
145 & O9.5V & 34673.7 & 11.62 &  4.23 & -3.23 &  4.62 \\
169 & B1.5V & 22387.2 & 13.90 &  4.51 & -2.27 &  3.37 \\
174 & B1.5V & 22387.2 & 12.55 &  4.46 & -2.27 &  3.89 \\
187 & B0.5V & 28183.8 & 13.24 &  5.40 & -2.84 &  4.22 \\
213 & B0Vp & 31622.8 & 11.95 &  4.20 & -3.14 &  4.38 \\
215 & B1V & 25409.7 & 12.97 &  4.03 & -2.58 &  3.68 \\
217 & O7III & 39902.5 & 10.22 &  4.42 & -3.59 &  5.41 \\
227 & O9V & 35975.0 & 11.47 &  4.73 & -3.31 &  4.92 \\
250 & B1V & 25409.7 & 12.88 &  4.08 & -2.58 &  3.73 \\
258 & O8V & 38459.2 & 11.10 &  4.49 & -3.43 &  5.04 \\
259 & B0.5V & 28183.8 & 11.42 &  3.83 & -2.84 &  4.32 \\
292 & B1V & 25409.7 & 13.08 &  5.40 & -2.58 &  4.18 \\
299 & O8V & 38459.2 & 10.84 &  4.61 & -3.43 &  5.19 \\
300 & B1?V & 25409.7 & 13.05 &  4.24 & -2.58 &  3.73 \\
304 & B5Ie & 13182.6 & 11.46 & 10.79 & -1.83 &  6.38 \\
317 & O8V & 38459.2 & 10.66 &  4.73 & -3.43 &  5.31 \\
339 & O8.5V & 37239.2 & 11.60 &  4.94 & -3.38 &  4.99 \\
358 & B2? & 19952.6 & 14.81 &  6.50 & -2.00 &  3.69 \\
376 & O8V & 38459.2 & 11.91 &  4.96 & -3.43 &  4.91 \\
378 & B0V & 31622.8 & 13.49 &  7.14 & -3.14 &  4.94 \\
390 & O8V & 38459.2 & 12.95 &  6.72 & -3.43 &  5.20 \\
395 & B1.5V & 22387.2 & 14.14 &  5.94 & -2.27 &  3.85 \\
403 & B1.5V & 22387.2 & 12.94 &  5.53 & -2.27 &  4.17 \\
417 & O4III & 48194.8 & 11.55 &  7.15 & -4.01 &  6.18 \\
421 & O9.5V & 34673.7 & 12.86 &  6.83 & -3.23 &  5.16 \\
425 & B0V & 31622.8 & 13.62 &  6.78 & -3.14 &  4.74 \\
426 & B0V & 31622.8 & 14.05 &  6.60 & -3.14 &  4.50 \\
429 & B0V & 31622.8 & 12.98 &  5.64 & -3.14 &  4.54 \\
431 & O5If & 44771.3 & 10.80 &  6.99 & -3.76 &  6.33 \\
448 & O6V & 43551.2 & 13.61 &  7.57 & -3.73 &  5.41 \\
455 & O8V & 38459.2 & 12.92 &  6.38 & -3.43 &  5.07 \\
457 & O3If & 50699.1 & 10.49 &  5.44 & -4.11 &  5.98 \\
462 & O6.5III & 41304.8 & 10.33 &  5.24 & -3.66 &  5.73 \\
465 & O5.5Ia & 43351.1 &  9.06 &  5.06 & -3.63 &  6.22 \\
467 & B1V & 25409.7 & 13.43 &  5.04 & -2.58 &  3.90 \\
470 & O9.5V & 34673.7 & 12.50 &  5.19 & -3.23 &  4.65 \\
473 & O8.5V & 37239.2 & 12.02 &  5.29 & -3.38 &  4.96 \\
480 & O7.5V & 39810.7 & 11.88 &  5.82 & -3.48 &  5.30 \\
483 & O5If & 44771.3 & 10.08 &  5.01 & -3.76 &  5.83 \\
485 & O8V & 38459.2 & 12.06 &  5.58 & -3.43 &  5.10 \\
507 & O8.5V & 37239.2 & 12.70 &  5.61 & -3.38 &  4.81 \\
515 & B1V & 25409.7 & 14.66 &  6.97 & -2.58 &  4.18 \\
516 & O5.5V & 44874.5 & 11.84 &  7.63 & -3.84 &  6.18 \\
531 & O8.5V & 37239.2 & 11.58 &  5.57 & -3.38 &  5.25 \\
534 & O7.5V & 39810.7 & 13.00 &  6.54 & -3.48 &  5.14 \\
555 & O8V & 38459.2 & 12.51 &  6.57 & -3.43 &  5.31 \\
556 & B1Ib & 26302.7 & 11.01 &  6.81 & -2.67 &  5.61 \\
575 & B1.5V & 22387.2 & 13.41 &  7.58 & -2.27 &  4.79 \\
588 & B0V & 31622.8 & 12.40 &  6.02 & -3.14 &  4.92 \\
601 & O9.5III & 32961.0 & 11.06 &  6.01 & -3.16 &  5.50 \\
605 & B0.5V & 28183.8 & 11.78 &  4.23 & -2.84 &  4.34 \\
611 & O7Vp & 41020.4 & 12.77 &  5.61 & -3.55 &  4.90 \\
632 & O9.5I & 31477.5 &  9.88 &  5.70 & -3.00 &  5.80 \\
642 & B1III & 23550.5 & 11.78 &  5.83 & -2.39 &  4.80 \\
646 & B1.5?V & 22387.2 & 13.34 &  5.13 & -2.27 &  3.84 \\
692 & B0V & 31622.8 & 13.61 &  5.69 & -3.14 &  4.31 \\
696 & O9.5V & 34673.7 & 12.32 &  5.85 & -3.23 &  4.99 \\
716 & O9V & 35975.0 & 13.50 &  6.10 & -3.31 &  4.65 \\
734 & O5If & 44771.3 & 10.04 &  5.54 & -3.76 &  6.06 \\
736 & O9V & 35975.0 & 12.79 &  5.46 & -3.31 &  4.68 \\
745 & O7V & 41020.4 & 11.91 &  5.30 & -3.55 &  5.12 \\
771 & O7V & 41020.4 & 12.06 &  7.00 & -3.55 &  5.73 \\
793 & B1.5III? & 22908.7 & 12.29 &  6.13 & -2.33 &  4.69 \\
S03 & O8.5III & 35727.3 & 10.22 &  6.01 & -3.30 &  5.92 \\
S73 & O8V & 38459.2 &  0.00 &  6.04 & -3.43 &  5.28 \\
A20 & O8II & 37153.5 &  0.00 &  7.77 & -3.37 &  5.88 \\
A23 & B0.7Ib & 26915.3 &  0.00 &  7.28 & -2.73 &  5.55 \\
A27 & B0Ia & 28840.3 &  0.00 &  7.06 & -2.88 &  5.75 \\
A29 & O9.7Iab & 30199.5 &  0.00 &  6.92 & -3.00 &  5.48 \\
A32 & O9.5IV & 34673.7 &  0.00 &  6.73 & -3.23 &  5.43 \\
A34 & B0.7Ib & 26915.3 &  0.00 &  5.96 & -2.73 &  5.21 \\
A36 & B0Ib & 28840.3 &  0.00 &  6.51 & -2.88 &  5.48 \\
A37 & O5V & 46131.8 &  0.00 &  6.77 & -3.95 &  5.51 \\
A39 & B2V & 19952.6 &  0.00 &  5.89 & -2.00 &  4.42 \\
A41 & O9.7II & 31622.8 &  0.00 &  5.97 & -3.14 &  5.29 \\
A42 & B0V & 31622.8 &  0.00 &  5.83 & -3.14 &  4.72 \\
A44 & B0.5IV & 28183.8 &  0.00 &  5.04 & -2.84 &  4.62 \\
A45 & B0.5V & 28183.8 &  0.00 &  4.64 & -2.84 &  4.53 \\
A46 & O7V & 41020.4 &  0.00 &  4.56 & -3.55 &  5.19 \\
\enddata
\end{deluxetable}

\section{DISCUSSION}

Comer\'{o}n et al.\ (2002) estimate that they have uncovered between 
90 and 100 O-type stars or closely related objects in the Cyg OB2 
association. Indeed, our results suggest the numerous OB candidates
they have identified are likely to be OB stars.   The fraction of O-stars 
in our sample of 14 newly identified OB stars is lower than they had 
predicted for the candidate sample (77\%), though if we include the 
B supergiants which {\sl used} to be O stars, we get better agreement 
(60\%) when one considers the sample size.

However, the high fraction of suspected non-members found in this
study of 14 new OB stars would suggest that many if not most of the
candidate OB stars identified by Comer\'{o}n et al.\ (2002) and
still yet to be confirmed will also be non-members.
The very extended spatial distribution of the CPR2002 sample, as
shown in Fig.\ 4, also supports this prediction.  The Cyg OB2 cluster 
may be more extensive and contain more O stars than previously thought, 
however it now seems less likely that 
a concentrated effort would yield over 100 O stars 
as predicted by Kn\"{o}dlseder (2000).  Based on this definition  
used by Kn\"{o}dlseder (2000), Cyg OB2 would not be classified 
as a super star cluster.  Naturally, spectroscopic observations of the 
remaining OB candidates listed by Comer\'{o}n et al.\ are needed to
resolve this debate.  Moreover, we can not rule out that deeper 
near-infrared investigations concentrated on the center of the the 
Cyg OB2 cluster region may reveal the underlying massive cluster 
the Kn\"{o}dlseder (2000) near-infrared study suggests.

\subsection{Does the Milky Way contain Super Star Clusters?}

The term ``super star cluster'' was first used to describe very
luminous young star clusters in nearby late type galaxies (Hodge 1961).
Early examples include 30 Dor in the Large Magellanic Cloud and NGC 330 
in the Small Magellanic Cloud, each harboring tens of thousands of stars.  
No such galactic counterparts, outside of the very old globular clusters, 
were known to exist in the galaxy.  Recently, the term has been used to 
describe quite a number of galactic clusters, first the Arches and 
Quintuplet clusters near the galactic center, and now Cygnus OB2 
(Kn\"{o}dlseder 2000) and Westerlund 1 
(Negueruela \& Clark 2003).  Based on the proximity of the last two, and 
assuming a similar distribution within the solar circle, Kn\"{o}dlseder 
et al. (2002) predict there may be as many as 100 similar clusters 
in the Galaxy.  Interestingly, evidence for such a large population 
of very massive clusters, is not unprecedented.
Extrapolating the cluster luminosity function of our 
galaxy, van den Bergh \& Lafontaine (1984) predicted a total of $\sim$ 
10$^2$ clusters with $M_V$ = -11, the absolute visual magnitude of 30 Dor!  
A continued extrapolation of our galaxy's cluster luminosity function 
predicts our galaxy to contain one cluster with $M_V$ = -12 (Larsen 2002).

Van den Bergh \& Lafontaine (1984) found it hard to believe clusters with
such mass could be contained within our galaxy.   Perhaps for this reason, 
it's been assumed that our galaxy's cluster luminosity function steepened 
over the range $-11 < M_V < -8$ (Larsen 2002).  However, there are many
factors contributing to massive clusters being hidden from view.  Most 
important is the fact that the most massive open clusters are also among the 
youngest (see Fig.\ 6 of Larsen \& Richtler 1999).   This is because
very massive and presumably more extended clusters are preferentially 
destroyed over lower mass clusters by dynamical friction in a relatively 
short timescales (10$^9$ yrs; de Grijs et al.\ 2003).  Recently, Portegies
Zwart et al.\ (2001; 2002) showed the exceedingly fast dynamical evolution
and fatal disruption of massive clusters formed in the inner 200 pc
of the galaxy, predicting their apparent loss of detection in just
5 million years.  They predict the central portion of our galaxy
could easily harbor as many as 50 clusters with properties similar
to the Arches or Quintuplet massive cluster systems.  Thus, our 
galaxy's most massive clusters are severely affected by 
line of sight (due to their close proximity to the galactic plane)
and local extinction (due to their youth) severely hindering our
efforts to find such objects. Given the rather low star formation
rate (SFR) seen at the sun's galactic radius, and the known
correlation between SFR and cluster luminosities
(Larsen 2002 and references therein), one would not expect to
find examples of very massive star clusters locally.  An absence of massive 
clusters in the well sampled, but small region out at the galactic
radius of our Sun can not be used as evidence for the absence of massive
clusters elsewhere.  Finally, the known absence of massive clusters 
in M31 while considering they might yet exist in the Milky Way is reasonable
considering M31's earlier Hubble type (Kennicutt \& Chu 1988).

{\subsection{How might galactic Super Star Clusters be found?}

Historically, radio surveys of the galactic plane have been successful to 
identify very young and massive galactic HII regions (Westerhout 1958), and 
to trace the structure of our galaxy (Georgelin \& Georgelin 1976).  A recent
series of papers by Blum and collaborators (Blum, Damineli, \& Conti 1999; 
Blum, Conti, \& Damineli 2000, Blum, Damineli, \& Conti 2001) have further 
investigated some of the largest of these radio identified HII regions.
Referred to as Giant HII regions (GHII), these HII regions have Lyman 
continuum fluxes more than 10 times that of Orion, or about 10 O stars.  
Their near-infrared studies were able to spectroscopically confirm OB stars 
in all three GHII regions.  However, emission line sources consistent with very 
massive young stellar objects were also found to lie among the brightest 
sources in each of the three clusters, a clear indication of the very
young age of clusters (not more than perhaps 1 million years old) found 
via radio surveys.  Curiously, the spectroscopic distances of all 
three of the GHII regions studied by Blum and collaborators turned out 
to be {\sl nearer} than the radio measurements had estimated.  This 
significantly reduced their intrinsic Lyman continuum flux, demoting two of
them from being true GHII regions to simply large HII regions. Most important,
their results indicate we are still biased to detecting HII regions on the 
near side of the galaxy, even when using radio surveys.  Only a few true 
GHII regions found via radio surveys are confirmed to lie on the far side
of the galaxy (such as W49A, see Conti \& Blum 2002).

Neither Cyg OB2 nor Westerlund 1 are easily detected in the radio.  This is
an important point to recognize.  Massive clusters older than a few million years 
will have been missed in galactic radio surveys.  How do we find our Galaxy's 
super star clusters?

Systematic infrared surveys to search for massive clusters with follow up 
near-infrared spectroscopic studies of their stellar members may be the
only method.  Sifting through deep near-infrared imaging surveys to search
for new massive clusters will be an important direction for galactic studies 
in the next decade.  Equally important will be determining the completeness 
limits for such objects, though it may only be possible within a few kpc from 
the Sun.  It is important to know if massive clusters found in the disk share 
any of the extreme qualities of the
very massive clusters found at the galactic center.  To do so means uncovering
and then investigating a significant number of disk super star clusters (if they
exist) to get a statistically meaningful result on the characteristics of
this new class of galactic objects.

The Cyg OB2 cluster serves as an important first calibration.  Because it has 
significant extinction,  Kn\"{o}dlseder et al. (2002) was forced to investigate, 
isolate and characterize the cluster using near-infrared methods.  However, the 
extinction is not so great that cross-checks with more traditional cluster 
studies, namely optical MK classification of many of its members, was 
ruled out.  Combining near-infrared and optical observations to study Cyg OB2 
has given us the best of both methods and resulted in a clearer description of 
the cluster and member properties.  The 100\% success rate, so far, of 
the CPR2002 study to select early-type stars from a dense field using very low 
resolution near-infrared measurements alone, gives great confidence in the 
ability to apply similar techniques to heavily 
reddened, near-infrared clusters now being uncovered. MK classification spectra 
are the most desirable, but for stars with A$_V > 8$ or so, the benefits of a 
more accurately determined optical spectral type are eclipsed by the need for 
a very large telescope to obtain such spectra.   For these stars, and this 
would include the remaining 31 OB candidate stars listed by CPR2002 which were
too extinguished for this study, stellar classification will need to be made at 
near-infrared wavelengths (Hanson et al.\ 1996).

\section{CONCLUSIONS}

New optical MK classification spectra have been obtained for 14 candidate OB stars from 
the Cyg OB2 cluster which were sifted and identified using near-infrared photometric
and low resolution spectroscopic measurements.  Our optical spectra confirm the 
early-type nature of these stars and lends support to Kn\"{o}dlseder's (2000) 
result that Cyg OB2 contains more early-type stars than previously thought.  However, 
about half of the OB stars appear to be non-members based on their evolutionary age.  
For this and other reasons, the present O-star count, still under 100, 
does not yet allow Cyg OB2 to be classified as a super star cluster,
as defined by Kn\"{o}dlseder et al. (2002). 

Our investigation of the Cyg OB2 cluster characteristics indicates that its distance
may be slightly closer than previous work quoted (DM = 11.2; Massey \& Thompson
1992) when the very young age, and thus under-luminous nature of these stars (as
compared to the more evolved clusters used in the derivation of dwarf star 
absolute magnitudes) is taken into consideration.  A revised distance of DM = 10.8
is predicted based on fitting 35 dwarf cluster members with spectral types
between O7.5 and B1 to a 2 x 10$^6$ year isochrone.   We also explored the newly 
presented effective temperature scale for dwarf O stars of Martins 
et al.\ (2002), using a similar fit of the dwarf population to 
a 2 x 10$^6$ year isochrone.  The new lower temperatures and thus lower absolute 
luminosities (particularly for mid-O dwarf stars) gives a very close 
value for the distance to Cyg OB2 (DM = 10.4).  This
lends some urgency to further observational testing and investigation of the 
new lower effective temperature scale for O dwarf stars, as it looks to be 
strongly supported by many independent theoretical groups at this time.  Observational
studies of additional massive clusters containing a significant sample of OB 
dwarf stars would be useful in this regard.

Our study has examined and confirmed the reliability of near-infrared 
broad band colors combined with low resolution survey spectra to accurately 
sift OB stars from a dense field population.  It would appear evidence for
very massive clusters in our galaxy is mounting, based both on direct observations
of fairly massive clusters near the Sun, as well as extrapolations of our galaxy's
cluster luminosity function.  Neither optical nor radio searches are likely 
to be successful in
locating these objects.  However, numerous near-infrared  clusters are currently
being uncovered with 2MASS. For extinction values A$_V > 8$, determination of 
spectral types for the stars in these clusters will need to be obtained using 
near-infrared classification techniques.  Great possibilities lie ahead for 
the discovery and precise study of very massive galactic OB clusters, uncovering 
very massive, possible, super star clusters currently hidden within the Galaxy.

\acknowledgements

We are grateful to Steward Observatory and The University of Arizona for 
their support of this program through their generous allowance of
telescope time. 
Phil Massey gave the author extensive, critical comments 
on the manuscript and, along with Nolan Walborn, performed an independent 
check of the MK spectral types for the stars in this study.  
The author continues to benefit from enlightening discussions with
Peter Conti. Finally, the referee provided 
insightful suggestions that greatly improved the quality of the manuscript.
The Digitized Sky Surveys were produced at the Space Telescope Science 
Institute under U.S. Government grant NAG W-2166. Near-infrared photometry
was obtained from The Two Micron All Sky Survey (2MASS), a joint project 
of the University of Massachusetts and the Infrared Processing and Analysis 
Center/California Institute of Technology, funded by the National Aeronautics 
and Space Administration and the National Science Foundation. This manuscript
is based upon work supported by the National Science Foundation under 
Grant AST-0094050 to the University of Cincinnati.

\end{document}